\documentclass[apjl]{emulateapj}
\usepackage{graphicx}
\usepackage{natbib}
\usepackage{amsmath}
\usepackage{multirow}
\usepackage{array}
\usepackage{subfigure}
\usepackage{morefloats}
\begin{document}

\title{A Tether-Assisted Space Launch System for Super-Earths}

\author{Alex R. Howe}
\affil{Department of Astronomy, University of Michigan}
\affil{arhowe@umich.edu}


\begin{abstract}

Super-Earths are one of the most common types of extrasolar planets currently known. Hundreds of these planets have been discovered by the $Kepler$ spacecraft and other surveys, with masses up to 10 $M_\oplus$ and consequently higher surface gravity than Earth. These planets would have greater escape velocities than Earth, reaching as high as 27 km s$^{-1}$ for the largest super-Earths. To launch a spacecraft from the surface of such a world with chemical rockets would be cost-prohibitive or nearly so, while another commonly proposed launch system, the space elevator, would lack the necessary tensile strength to support its weight. However, we find that a hybrid launch system combining chemical rockets and a space-based momentum-exchange tether could reduce the $\Delta v$ to be provided by chemical rockets to reach escape velocity by 40\%, bringing it back into the realm of feasibility. Such a system could also function on Earth with considerably less exotic materials.

\end{abstract}

\maketitle


\section{Introduction}

The most common type of extrasolar planets currently known are the super-Earths, those planets between Earth and Neptune in size. Recently, this population has been divded into two subpopulations: those still designated super-Earths, which have a rock-iron composition similar to Earth, but are larger, and the mini-Neptunes, which have a gaseous envelope similar to Neptune, but are smaller \citep{Fulton}. Rocky super-Earths have been observed with masses up to 9.7 $M_\oplus$ with a radius of 1.89 $R_\oplus$ for Kepler-20b \citep{Buchhave}, and the densest super-Earths likely to exist may be estimated at 10 $M_\oplus$ and 1.7 $R_\oplus$ \citep{Hippke}.

Some super-Earths may be habitable, and indeed, some of their properties suggest that they may be even more amenable to life than Earth is \citep{Heller}. So the possibility of intelligent life arising on such a world is worth considering. However, an exo-civilization developing on a super-Earth would find itself at a disadvantage in the area of space travel because of the high escape velocity (and orbital velocity) of super-Earths. The densest super-Earths would have a surface gravity of 3.46$g$ and an escape velocity of $\sim$27.1 km s$^{-1}$, compared with $\sim$11.2 km s$^{-1}$ for Earth. Because the amount of fuel required by chemical rockets increases exponentially as a function of the change in velocity ($\Delta v$), achieving escape velocity from a super-Earth would require many times more fuel than on Earth. \cite{Hippke} finds that launching a spacecraft from the surface of Kepler-20b would require 104 times as much fuel as launching from Earth, requiring proportionately larger chemical rockets, which would be prohibitively expensive to space programs like those on Earth.

In this paper, we explore the possibility of launching a spacecraft from the surface of a super-Earth when chemical rockets are supplemented by a momentum-exchange tether, which can greatly increase the craft's speed relative to the planet at no fuel cost. We find that such a tether system could reduce the $\Delta v$ demanded of chemical rockets to reach escape velocity by 40\%, bringing them back into the range of rockets launched from Earth. In Section \ref{Earth}, we also examine the applicability of such a system to Earth-based launches.


\section{Limits of Chemical Rockets}

All orbital launch systems on Earth currently require chemical rockets to overcome Earth's gravity, as this requires not just energy, but large amounts of thrust that cannot be achieved with a more efficient engine such as an ion thruster. Unfortunately, chemical rockets are very fuel-inefficient because of their low specific impulse, $I_{sp}$, which is roughly $\sim$350-450 s for hydrogen/oxygen fuel \citep{Hippke}. The total $\Delta v$ imparted by a chemcial rocket is given by the Tsiolkovsky rocket equation:
\begin{equation}
\Delta v = v_{ex}\ln{\frac{m_0}{m_f}},
\end{equation}
where $v_{ex}$ is the exhaust velocity of the fuel, $v_{ex}\approx g_0 I_{sp}$, $m_0$ is the initial mass of the spacecraft with fuel, and $m_f$ is the final mass of the spacecraft without fuel. Thus, the fuel cost increases exponentially as the $\Delta v$ demands of the launch increase.

For a launch from a large super-Earth to the escape velocity of 27.1 km s$^{-1}$, given an oxygen/hydrogen engine with a specific impulse of 400 s, the required mass ratio of fuel to empty weight is 1,006. In practice, a multi-stage rocket may improve this, but it would still be much greater than the largest rocket ever flown on Earth, the \textit{Saturn V}, which had a mass ratio of 68 with a launch weight of 2,970 tonnes. Thus, even the largest rocket yet constructed could lift only a small payload to escape velocity$-$less than 3 tonnes including fuel tanks and structural supports, if it could be made structurally strong enough with such a large mass ratio.

A great deal depends on the specific impulse of the engine. For $I_{sp} = 350$ s, the mass ratio evaluates to 2,700, while for $I_{sp} = 450$ s, the mass ratio evaluates to 466. All of these values are in the possible range for hydrogen/oxygen fuel. Nonetheless, all of them are much larger the the mass ratios of real rockets, so while it may be possible to build rockets this large, the cost would most likely be prohibitive.

The performance of a chemical rocket can be improved by using a fuel with a higher specific impulse. The most powerful chemical fuel ever tested was a lithium, fluorine, and hydrogen mixture with $I_{sp} = 542$ s \citep{Arbit}. This fuel would decrease the mass ratio of a rocket launched from a large super-Earth to 164. However, the technical and supply considerations of a fuel containing large amounts of fluorine and lithium mean that this would not be a great improvement in terms of cost. Thus, it is unlikely that the technology of chemical rockets alone can support a space program on a super-Earth.

Another option is not to attempt to reach escape velocity, but instead to reach a low orbit. Once in orbit, a spacecraft can move away from the planet slowly using a much more efficient ion thruster, which requires negligible fuel compared with chemical rockets \citep{Jahn}. The orbital velocity in low orbit is given by $v_{orb} \approx v_{esc}/\sqrt{2}$, so for a large super-Earth, the orbital velocity is 19.2 km s$^{-1}$. For a chemical rocket with $I_{sp} = 400$ s, a mass ratio of 134 is needed to reach low orbit. This is still more difficult than launches from Earth, but readily achievable with current technology.


\section{Momentum Exchange Tethers}

One proposed alternative to chemical rockets is a space-elevator, which extends a tower or cable from the planet's surface to beyond the synchronous orbit, allowing an electric vehicle to climb and reach orbit or even escape velocity it at no fuel cost. However, construting a cable that can hang under such stress would require materials with immense tensile strength that may not be realizable even for Earth, and would almost certainly not be able to withstand the stronger gravity of a super-Earth \citep{Pugno}.

A momentum exchange tether is a cable system that does not have a fixed end, but instead rotates in orbit around a planet \citep{Moravec}. In such a system, a spacecraft docks at one end of the tether and rotates around to the opposite position, where it is released, accelerating it relative to the planet by the exchange of angular momentum, again without any fuel cost. While the tether loses angular momentum in the process, causing its orbit to decay, this angular momentum can be recovered in several ways at very little cost, including a momentum exchange in the other direction with an incoming payload to the planet, a fuel-efficient ion thruster to raise its orbit, or by electrodynamic propulsion, which generates thrust from the planet's magnetic field \citep{Katz}.

The most efficient form of momentum exchange tether, sometimes called a ``skyhook'', would be a tether that rotates at the planet's orbital velocity such that its ends trace out an epicycloidal path around the planet. In the frame of reference of the planet's surface, the end of the tether would descend vertically and come to a momentary stop at a relatively low altitude, at which point a suborbital spacecraft could dock with it, before ascending back to orbit. The tether would also be much smaller than a space elevator$-$possibly as short as a few hundred kilometers in length compared with 100,000 km for a space elevator. However, rotating a tether to even the low-Earth orbit speed of 7.8 km s$^{-1}$ would require a similar tensile strength to a space elevator, and such a system would not be able to function on a super-Earth. Nonetheless, a momentum-exchange tether system would still reduce the $\Delta v$ needed for chemical rockets to reach escape velocity by twice the rotational speed of the tether, even if it does not rotate at the orbital velocity.

The functional limit of a momentum exchange tether is given by a \textit{characteristic speed} of the material, $v_c$, which is the maximum speed at which an untapered cable can rotate without breaking. (Tapering the tether would add additional strength, but for this paper, we use the untapered strength and assume any tapering is used to provide a safety margin.) The characteristic speed is given by:
\begin{equation}
v_c = \sqrt{\frac{2\sigma}{\rho}},
\end{equation}
where $\sigma$ is the tensile strength of the tether material, and $\rho$ is its density. The strongest known material for constructing such a tether is carbon nanotubes. \cite{Pugno} estimates the tensile strength of a macroscopic cable made out of carbon nanotubes at $\sigma = 22$ GPa and $\rho = 1.5$ g cm$^{-3}$, which result in $v_c \approx 5.4$ km s$^{-1}$.


\section{Launch System}

A hybrid launch system would involve a suborbital chemical rocket to reach a momentum-exchange tether in low orbit, where the $\Delta v$ requirement to reach orbit is reduced by the rotational speed of the tether. The spacecraft would dock onto the end of the tether and rotate around until it is released at the top of the arc at a higher speed. At that point, additional chemical rockets would be used to reach escape velocity (although an ion thruster would also work at this point). With a rotational speed of 5.4 km s$^{-1}$ for a carbon nanotube tether orbiting a large super-Earth, the required $\Delta v$ for chemical rockets would be reduced from 27.1 km s$^{-1}$ to 16.3 km s$^{-1}$. For this $\Delta v$, the mass ratio of an average hydrogen-fuelled rocket evaluates to 64, similar to the \textit{Saturn V}, so this system would be in line with launches from Earth.

Building a momentum-exchange tether without an existing tether-assisted system in place would be slightly more complicated. However, for this purpose, a small ``seed'' tether could be launched into orbit conventionally by chemical rockets, which can reach low orbit without assistance. After this, robotic climbers could be launched on suborbital rockets to dock onto the tether and add more strands, thickening it and increasing its maximum payload. Thus only a single conventional launch would be needed to build a tether-assisted launch system.

Once a momentum-exchange tether is in place and built up enough to carry large payloads, launching to space from a super-Earth would be much easier, albeit still more difficult than on Earth. The $\Delta v$ required to reach the tether with chemical rockets would be 13.8 km s$^{-1}$, and the docking would have to be achieved in the span of a few seconds in which the spacecraft's speed nearly matches the tether's. From here, a climber could crawl to the midpoint of the tether, at which point it would be in low orbit, or the spacecraft could rotate around and be released at the top of the tether's arc. From this point, an acceleration of a further 2.5 km s$^{-1}$ would be needed to reach escape velocity, but interestingly, this is less than the 3.4 km s$^{-1}$ needed to reach escape velocity from low-Earth orbit without tether assistance.


\section{Application to Earth}
\label{Earth}

A momentum exchange tether is an even greater advantage for spaceflight on less massive planets because the lower speeds involved make it easier to reach or at least approach the orbital speed of the planet. For Earth, even with our current level of technology, a momentum exchange tether in Earth orbit would greatly reduce the fuel cost for launches to orbit and elsewhere in the solar system. Aramid-type fibers such as Kevlar are strong enough to achieve characteristic velocities approaching 3.0 km s$^{-1}$. Thus, a momentum exchange tether could reduce the $\Delta v$ required for chemical rockets to reach Earth orbit from 7.8 km s$^{-1}$ to 4.8 km s$^{-1}$, and the $\Delta v$ to reach escape velocity from 11.2 km s$^{-1}$ to 5.2 km s$^{-1}$.

A $\Delta v$ of 5.2 km s$^{-1}$ with oxygen-hydrogen fuel could allow a mass ratio as low a 3.25 for an efficient enough engine and consequently a much smaller and less powerful rocket. With such a system, even a suborbital rocket could reach interplanetary space, making it a great improvement on current rocket technology.


\section{Conclusion}

The observed range of super-Earths extends up to $\sim$10 $M_\oplus$, and the high masses and surface gravities of these planets would present a serious challenge to space travel. For the largest super-Earths, most proposed space launch systems (short of nuclear propulsion) would range from infeasible to impossible for reaching escape velocity, although reaching low orbit would be possible. However, a hybrid system combining chemical rockets and a momentu-exchange tether could put interplanetary travel within reach for a civilization on even the largest terrestrial planets, bridging the gap by combining the maximum feasible contributions of each system to reach a sufficient velocity to escape the planet's gravity well. Combining chemical rockets with ion thrusters would also work, but would present less of an advantage.

The same method could also reduce the cost of space launches on Earth much more easily than with a space elevator. While a space elevator requires roughly 100,000 km of cable and a strength at the limit of what is physically possibly, a momentum exchange tether in low-Earth orbit would require less than 1,000 km of cable and could boost suborbital launches to escape velocity using currently available materials.


\bibliographystyle{apj}
\bibliography{apj-jour,refs}


\end{document}